\documentclass[aps,prl,superscriptaddress,twocolumn]{revtex4}

\usepackage[dvips]{graphicx}
\usepackage{amsmath}
\usepackage{amssymb}
\usepackage{bm}

\begin{document}

\title{Renormalization group analysis of the $M$-$p$-spin glass model
with $p=3$ and $M=3$}

\author{Joonhyun Yeo}
\affiliation{Division of Quantum Phases and Devices,
School of Physics, Konkuk University, Seoul 143-701, Korea}
\author{M.\ A.\ Moore}
\affiliation{School of Physics and Astronomy, University of Manchester,
Manchester M13 9PL, UK}

\date{\today}

\begin{abstract}
We  study an $M$-$p$-spin  spin glass  model with  $p=3$ and  $M=3$ in
three  dimensions  using  the  Migdal-Kadanoff  renormalization  group
approximation (MKA). In this version  of the $p$-spin model, there are
three ($M=3$) Ising spins on  each site. At mean-field level,
this model  is known to  have two transitions; a  dynamical transition
and  a  thermodynamic  one  at  a  lower  temperature.  The  dynamical
transition  is similar  to  the mode-coupling  transition in  glasses,
while the thermodynamic transition  possibly describes what happens at
the Kauzmann temperature.  We find that all the  coupling constants in
the model  flow under  the MKA to  the high-temperature  sink implying
that the mean-field features disappear in three dimensions  and that there is
no transition  in this  model. The behavior  of the  coupling constant
flow  is qualitatively similar  to that  of the  model with  $p=3$ and
$M=2$,  for which  only a  single transition  is predicted  at  the mean-field
level.   We conclude  that for  $p$-spin models  in  three dimensions,
fluctuation effects completely remove all traces of their mean-field behavior.
\end{abstract}



\maketitle


Understanding the dynamics and thermodynamics of supercooled liquids
and structural glasses is one of the most challenging problems in 
condensed matter physics. Theoretical advances have been made 
using an idea obtained from certain mean-field spin glass models \cite{KTW}. 
One example is the infinite-range $p$-spin  spin   glass  model \cite{KTW},
which shows two phase transitions. There is a  
dynamical transition at a  temperature $T_d$ below which an ergodicity
breaking  occurs. The connection to structural glasses
comes from the similarity of the dynamical equations of 
these spin glass models near $T_d$ to those  in  the mode-coupling  theory of  the
structural  glass transition \cite{MCT}.  Below the dynamical transition, a
thermodynamic  transition takes  place  at $T_0$ with the low temperature
phase described by the one-step replica symmetry breaking (RSB).  
This  transition  has been  associated with  a
possible transition at the Kauzmann temperature $T_K$ \cite{Kauzmann}
in structural glasses.
An important question is then how this mean-field 
picture gets changed in finite dimensions.
Beyond the mean-field level, the dynamical transition at $T_d$ is expected to disappear
because of activation processes over finite energy barriers separating 
metastable states. 
However,  the  thermodynamic glass  transition
$T_0$   could  still   exist  in   principle  in   finite  dimensional
systems. In this paper, we use the Migdal-Kadanoff (MK) real space renormalization 
group (RG) method to investigate this issue.  

The original $p$-spin glass model mentioned above, however, is inconvenient to study on 
a simple hypercubic lattice.  In Ref.~\cite{PPR}, a new class of $p$-spin glass 
models called the $M$-$p$-spin glass model has been proposed. 
These models consist of Ising spins on a hypercubic
lattice with  nearest-neighbor $p$-body interactions. The new feature of these models
is that there are $M$ Ising spins on each site. 
As the value of the parameters $M$ and $p$ change, 
these models exhibit two distinct types of behavior at mean-field (MF) level.
For example, when $p=3$ and $M=2$, the model at mean-field level
has a single continuous transition to a full RSB state. 
On the other hand, for $p=3$ and $M=3$, one has the two transitions
described above. One can actually classify all these models for general $p$ and $M$, 
into the two mean-field transition types \cite{mpspin}. 
Therefore the $M$-$p$-spin glass model is an ideal testbed for
studying how thermal fluctuations change mean-field results. 
The $p=3$, $M=2$ model has been studied by Monte Carlo simulations in four dimensions \cite{PPR}
and the results were interpreted as evidence for  a single  continuous transition. 
The Migdal-Kadanoff RG has been applied \cite{DBM,MD} to this model, using versions to mimic both three and four dimensions,
and it was shown that the coupling constants flow to the high-temperature sink
implying that there is no transition in both these dimensions.
This model has been shown \cite{MD} 
to be in the same universality class as the Ising spin glass in a field,
which is consistent with the absence of a phase transition in less than six dimensions \cite{AT}.
The correlation length
obtained from the Migdal-Kadanoff RG grows quite rapidly with decreasing temperature to a value of order $30$ lattice spacings. Thus in a numerical simulation of a finite system whose linear extent  is of order $10$ lattice spacings, it is therefore very likely that the rapidly growing correlation length seen would appear to be implying there is a transition \cite{PPR}, but this would just be an artifact of finite size effects.
The non-mean-field regime of the $p=3$ and $M=2$ model has also been simulated using
a one-dimensional long-ranged Ising spin glass where the couplings decrease with their separation $r_{ij}$ as $\sim 1/r_{ij}^{\sigma}$. By varying the exponent $\sigma$  
one can effectively tune the one-dimensional system to have similarities to the short-range problem in  
spatial dimension $d$ \cite{Larson}.
The mean-field transition was shown to be destroyed at values of $\sigma$ which correspond to low values of $d$.

It is a purpose of this paper to find whether the mean-field transitions in 
the $p=3$ and $M=3$ model, which are
relevant to discussions on structural glasses, survive 
to three dimensions using the Migdal-Kadanoff real space RG. The MKA is an approximate 
RG procedure which works best in low-dimensions $d$ such as  two and three where it gives excellent results for the Ising (i.e. $p=2$) spin glass.

The $M$-$p$-spin model we study in this paper is characterized by 
$M$ Ising spins $\sigma^{(\alpha)}_i$, $\alpha=1,2,\cdots,M$ on each site $i$
of a hypercubic lattice.
The Hamiltonian is given in terms of
products of $p$ spins chosen from the spins in a pair of nearest neighbor sites. 
In this paper, we focus on the $p=3$ case and 
the Hamiltonian of the 3-spin glass model is given by
\begin{eqnarray}
\mathcal{H}&=&-\sum_{\langle ij\rangle}\sum^M_{\alpha<\beta}\sum^M_{\gamma}\Big( 
J^{(\alpha\beta)\gamma}_{ij} \sigma^{(\alpha)}_i \sigma^{(\beta)}_i 
\sigma^{(\gamma)}_j \nonumber\\
&&\quad\quad\quad\quad\quad\quad +
J^{\gamma(\alpha\beta)}_{ij} \sigma^{(\gamma)}_i \sigma^{(\alpha)}_j 
\sigma^{(\beta)}_j\Big),
\end{eqnarray}
where the notation $\langle ij\rangle$ means that the sum is over all nearest neighbor pairs $i$ and $j$. 
Note that the number of different coupling constants, $J^{(\alpha\beta)\gamma}_{ij}$ 
and $J^{\gamma(\alpha\beta)}_{ij}$ for given $\langle ij\rangle$ 
is just $2M\binom{M}{2}=M^2(M-1)$.
All these couplings are chosen independently from a Gaussian distribution with zero mean 
and width $J$. 
Therefore for the 3-spin glass models, there are 4 and 18 independent couplings for $M=2$
and $M=3$, respectively.

The above $M$-$p$-spin model can be put into a field theoretical framework.
A standard way to do this is to use the Hubbard-Stranotovich transformation 
on the replicated partition function and then trace over the spins. 
The resulting field theory associated with this model is the following
Ginzburg-Landau-Wilson Hamiltonian 
\begin{eqnarray}
\mathcal{H}_{\mathrm{GLW}}&=&\int d^d\mathbf{r} \Bigg\{
\frac{1}{2}\sum_{a<b}[\nabla q_{ab}(\mathbf{r})]^2+\frac{t}{2}
\sum_{a<b}q^2_{ab}(\mathbf{r}) \nonumber \\
&&\quad\quad-\frac{w_1}{6}\mathrm{Tr} q^3(\mathbf{r}) 
-\frac{w_2}{3}\sum_{a<b} q^3_{ab}(\mathbf{r}) \Bigg\} ,
\label{GLW}
\end{eqnarray}
where $q_{ab}(\mathbf{r})$ is the order parameter and $a$ and $b$ are replica indices
running from 1 to $n$ with $n\to 0$.
At mean field level, this model is known \cite{Gross} to show very different behavior
depending on the value of $R=w_2/w_1$. 
When $R>1$, there are two transitions at the mean field level as described above; 
a dynamical transition
at some temperature $T_d$ and a thermodynamic transition at a lower temperature
$T_0$ to a state with one-step replica symmetry breaking.
The two transitions have been discussed extensively in connection with what happens 
in structural glasses near 
the mode-coupling and the Kauzmann temperatures, respectively.  
For the case where $R<1$, only a single transition 
to a state with full RSB is expected at mean-field level. In Ref.~\cite{mpspin}, the ratio $R$ was evaluated for 
the $M$-$p$-spin model for general values of $M$ and $p$. The cases we are
interested in this paper, namely $p=3, M=2$ and $p=3, M=3$, correspond to 
$R=3(1-1/\sqrt{2})\simeq 0.879$ and $R=2$, respectively. Therefore the two models
show very different mean-field behavior. It is the purpose of this paper to investigate
how thermal fluctuations affect the two kinds of mean-field behavior in three dimensions.

To do this we apply the Migdal-Kadanoff real space RG to these models.
We  follow closely  the approximate  bond moving  scheme  described in
detail  in Ref.~\cite{DBM}.   For a  three dimensional  cubic lattice,
four bonds are put together to form a new bond.  The coupling constant
of this new bond is just given by the sum of the coupling constants of
the four  bonds. We can  obtain a coarse  grained lattice by  taking a
trace  over  the spins  at  the site  connecting  two  new bonds.  The
decimation procedure can  be continued $n$ times for  a system of size
$L=2^n$.  Although  we  initially   start  from  only  the  three-spin
couplings   $K^{(\alpha\beta)\gamma}\equiv  J^{(\alpha\beta)\gamma}/T$
and   $K^{\gamma(\alpha\beta)}\equiv  J^{\gamma(\alpha\beta)}/T$ drawn from a Gaussian distribution with
standard deviation (width) $K\equiv  J/T$ at  temperature $T$, the  decimation procedure
generates additional types of couplings as  well as on-site  ``field'' terms
which  involve the  spins only  on one  site.  In  the Migdal-Kadanoff
analysis, we have to keep track of all these couplings.

\begin{figure}
\includegraphics[width=0.475\textwidth]{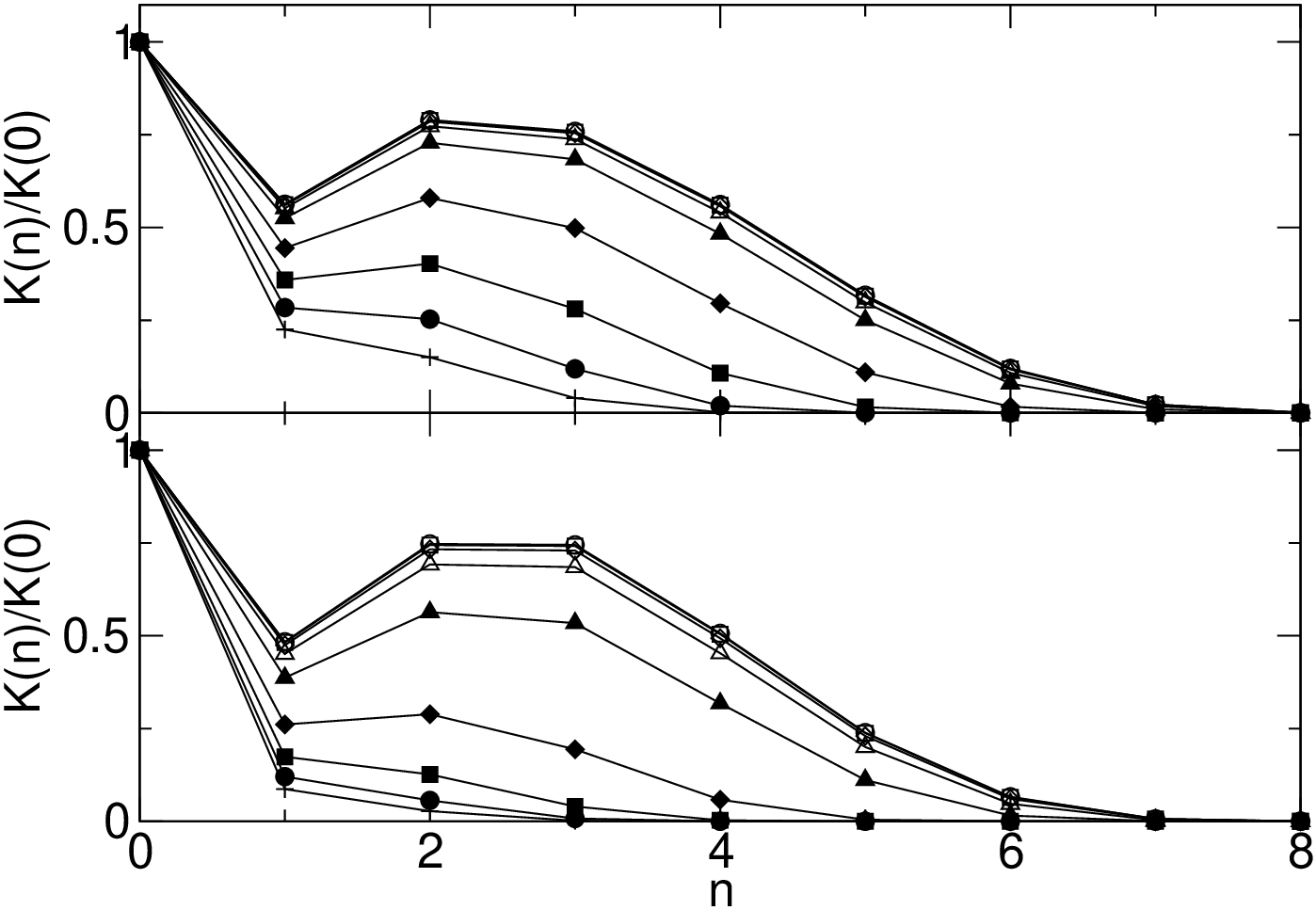}%
\caption{RG flow of the three-spin coupling strength $K(n)/K(0)$ normalized with 
its initial value at iteration step $n$. The upper panel is for $p=3$ and $M=3$,
and the lower one for $p=3$, $M=2$. One each panel, the curves correspond from bottom to top to 
temperatures $T/J=10, 8, 6, 4, 2, 1, 0.5, 0.25$ and 0.1.}
\label{three_spin}
\end{figure}

For the $p=3$, $M=2$ case, on each bond there are initially four three-spin couplings.
In addition to those, we have to consider
four two-spin couplings and one coupling that involves all the four spins
on the nearest-neighbor sites. 
There are also four field terms on the two nearest neighbor sites
and two on-site terms involving two spins on one site. If we include a constant 
contribution to the Hamiltonian, the renormalized Hamiltonian carries a total
of 16 coupling constants. For the $p=3$, $M=3$ case, on the other hand, each bond
is connected to 6 spins, 3 spins on each site. 
There are couplings involving 2, 3, 4, 5 and 6 spins
from a pair of nearest neighbor sites, whose standard deviations are denoted
by $C$, $K$, $D$, $E$ and $F$, respectively. The numbers of independent coupling constants
connecting 2, 3, 4, 5 and 6 spins are given by 9, 18, 15, 6 and 1, respectively.
There are also on-site terms. The numbers of on-site fields involving
1, 2 and 3 spins are 6, 6 and 2, respectively. Including a constant term,
we have to keep track of a total 64 couplings in the case of $p=3$ and $M=3$.

\begin{figure}
\includegraphics[width=0.45\textwidth]{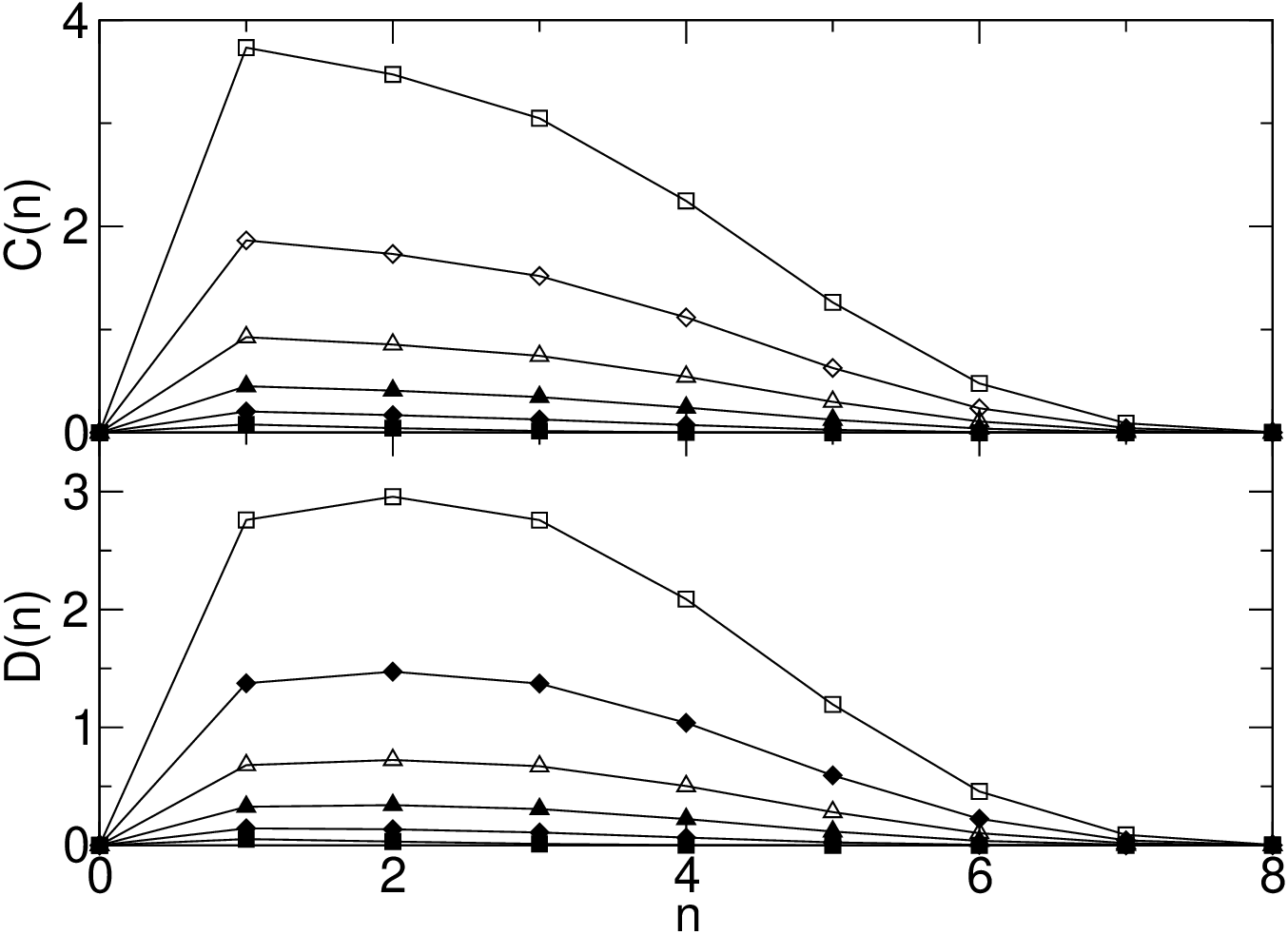}%
\caption{RG flow of the two-spin coupling $C(n)$ and 
the four-spin coupling $D(n)$ of the $p=3$, $M=3$ model at the $n$-th iteration step. 
The curves correspond from bottom to top to 
temperatures $T/J=8, 4, 2, 1, 0.5$ and 0.25.}
\label{two_four_spin}
\end{figure}

A key step in the calculation is setting up and solving the recursion relation
resulting from the decimation procedure.
For $p=3$ and $M=3$, 
this amounts to solving a system of 64 simultaneous equations for a new set of couplings,
which requires an inversion of a constant $64\times 64$ matrix. This only needs to  be
done once beforehand  and not for every iteration. In the actual numerical calculations, $10^6$ bonds are prepared.
On each bond 18 three-spin couplings, $K^{(\alpha\beta)\gamma}$ 
and $K^{\gamma(\alpha\beta)}$ are chosen independently from the Gaussian 
distribution with zero mean and width $K(0)=J/T$ at temperature $T$. 
All the other couplings including the fields are set to zero initially. We then randomly
choose two sets of 4 bonds to form two new bonds. Using the recursion relation,
we obtain a set of 64 renormalized couplings. This procedure is continued until 
we get $10^6$ new bonds, which completes the first iteration. As we iterate
the same procedure, we can obtain the flow of 64 couplings. At each step of the 
iteration, we evaluate the width of each distribution. We note that
there is a certain freedom in where to move on-site fields when the four bonds are put together.
The main results do not depend on the method chosen. 
We follow the prescription given in Ref.~\cite{DBM}: when three bonds are moved to 
combine with a bond, the fields on the three bonds are placed at the
site that is to be traced over.

The RG flow of the width of the three-spin coupling $K(n)$ at the $n$-th
iteration normalized with the initial value $K(0)=J/T$ is shown 
in Fig.~\ref{three_spin} for the 3-spin glass model
with $M=3$ and $M=2$, respectively. Somewhat surprisingly, both figures show essentially
the same features. At all temperatures, the couplings decrease to zero in the long
length scale limit indicating an RG flow to the high temperature sink.
This implies that there is no finite temperature phase transition 
in these models in contrast to the corresponding mean-field pictures. 
At high temperatures, the coupling simply decreases to zero. As
the temperature is reduced, however, the coupling increases during the first few
iterations reaching a maximum at the second or third iteration, but  eventually
decaying to zero. At very low temperatures $T/J\lesssim 0.25$, the flow patterns
of $K(n)/K(0)$ reach an asymptotic form. 
The only difference between $M=3$ and $M=2$ cases is that the couplings
in the $M=3$ model decay to zero more slowly as the iteration proceeds.
We can deduce from this that the correlation length in the $M=3$ model is longer
than that of the $M=2$ model.

\begin{figure}
\includegraphics[width=0.4\textwidth]{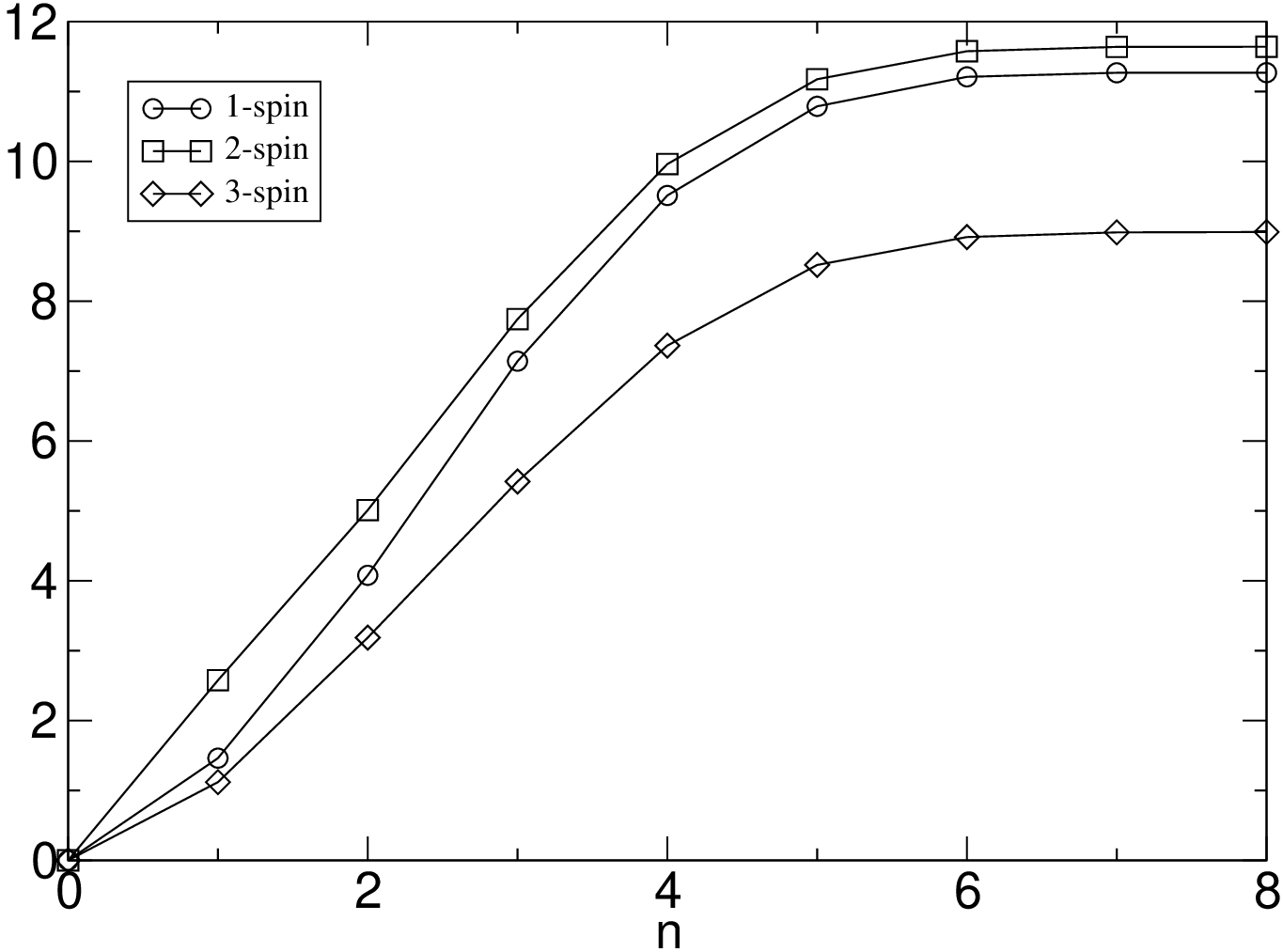}%
\caption{RG flow of on-site fields at temperature $T/J=0.5$ for $p=3$ and $M=3$. 
Circles, squares and diamonds correspond to the one-spin, two-spin and three-spin one-site terms, respectively. }
\label{field}
\end{figure}

The other couplings for the $p=3$ and $M=3$ model
also flow to zero as can be seen from Fig.~\ref{two_four_spin}.
The upper panel shows the RG flow of the width $C(n)$ of the two-spin couplings 
for various temperatures, which reaches a maximum at the first iteration and 
then decreases to zero as the iteration proceeds.  
The four-spin coupling strength $D(n)$ also shows 
the same behavior except that it has the largest values at the second iteration
step. The five- and six-spin couplings (not shown here) also exhibit essentially the same
behavior. The on-site fields, however, do not decrease but increase and saturate to 
a constant value in the long length scale limit as shown in Fig.~\ref{field}.  
The same behavior was observed 
in the Migdal-Kadanoff analysis of the $p=3$, $M=2$ model. Therefore, the $p$-spin glass model
for both $M=2$ and $M=3$ cases correspond on large length scales 
to a model with random fields acting on spins on decoupled sites.

\begin{figure}
\includegraphics[width=0.4\textwidth]{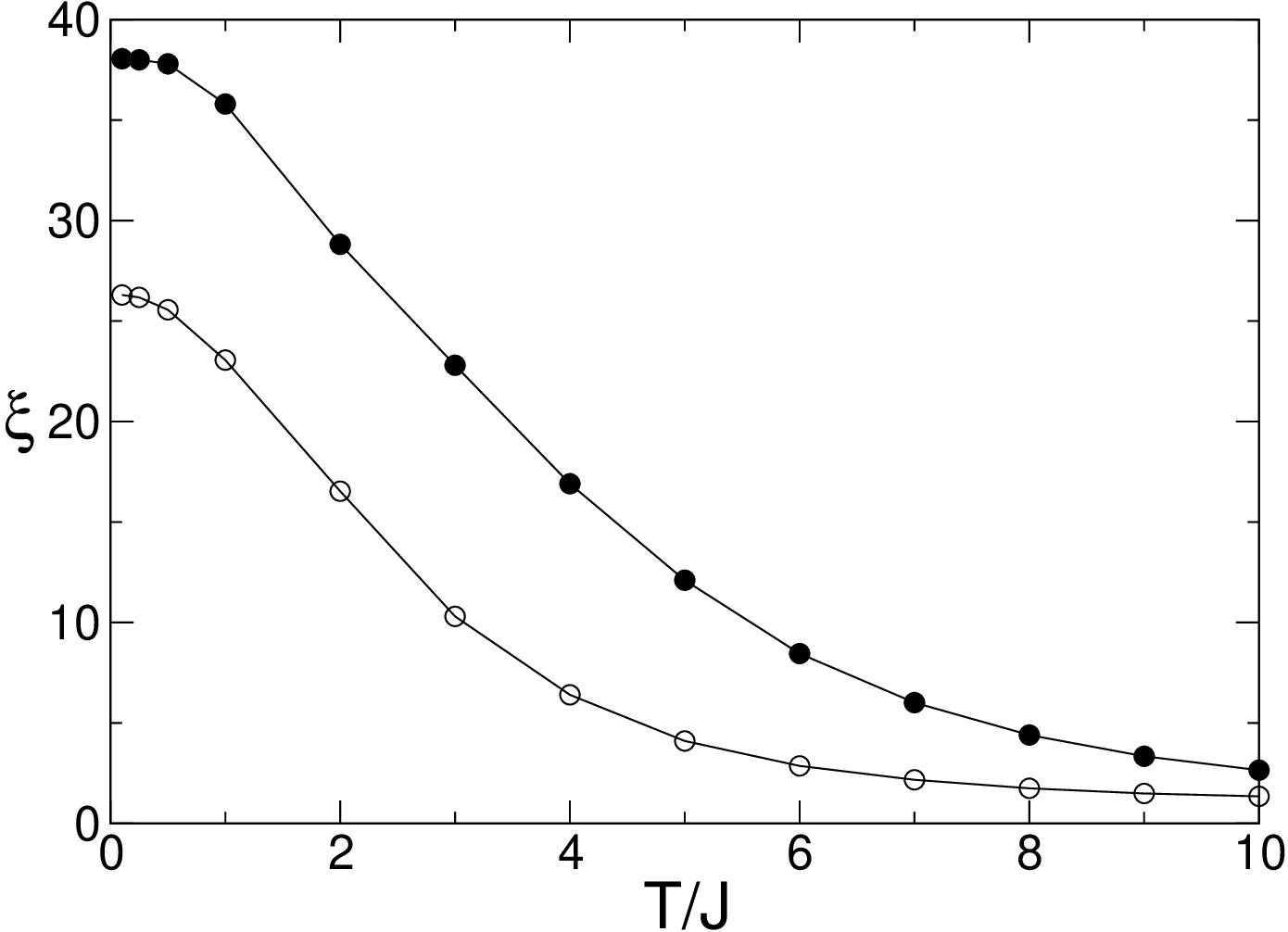}%
\caption{Correlation length $\xi$ as a function of 
temperature $T/J$ for the $p=3, M=3$ model (filled circles) 
and the $p=3, M=2$ model (open circles).}
\label{correlation}
\end{figure}

We now study how the standard deviations of couplings behave as a function of
the length scale $L=2^n$. The standard deviation $K$ of the three-spin couplings shown in 
Fig.~\ref{three_spin} decays as $\exp(-L/\xi(T))$, where $\xi(T)$ at temperature $T$
can be interpreted as the correlation length. In Fig.~\ref{correlation},
the correlation lengths of the two 3-spin glass models with $M=2$ and $M=3$ are shown. 
The correlation length 
for the $p=3$, $M=2$ model has been already obtained in Ref.~\cite{MD}.  
The $M=3$ model has longer correlation length  than the $M=2$ model at any given temperature. 
The correlation lengths of the two models basically show the same temperature
dependence. They increase quite rapidly as the temperature is lowered and then 
saturate  at  large values  in the zero temperature limit.

In  summary,  we  have  investigated  the RG  flows  of  the  coupling
constants  in the  $M$-$p$-spin glass  model with  $p=3$ and  $M=3$ in
three  dimensions using  the Migdal-Kadanoff  approximation.   All the
coupling  constants flow to  the high  temperature sink  implying that
there is  no finite  temperature phase transition  in this  model. The
correlation  length  increases quite  rapidly  as  the temperature  is
lowered  and  saturates  in  the  zero temperature  limit.  All  these
features are very similar to those of its $M=2$ counterpart. The only difference is
that the  $M=3$ model  has longer correlation  lengths. This  is quite
remarkable in  that the two models show  completely different behavior
at the  mean-field level. Therefore  we conclude that in  the $3$-spin
spin glass models, thermal fluctuations in three dimensions completely
destroy the transitions predicted  by mean-field theory: whether $R>1$
or  is less  than $1$  results in  only a  quantitative change  in the
magnitude of the correlation length.

What  stimulated  this  investigation  was the  renormalization  group
analysis by  Cammarota et al. \cite{CBTT} of  a replicated Hamiltonian
similar  to $\mathcal{H}_{\mathrm{GLW}}$ in  Eq. (\ref{GLW}).  Like us
they  used a  MK approximation  appropriate to  three  dimensions, but
instead  found a transition  which they  identified as  the transition
expected  in  random first-order  transition  theory  \cite{KTW} at  a
temperature  they identified as  $T_K$. We  believe the  difference in
results stems  from their  decimation step in  the MK procedure.  In our
work   it   is   carried   out   exactly.  In   the   calculation   of
Ref.  \cite{CBTT},   however,  it  was  done   by  a  steepest-descent
approximation to an  integral over $n \times n$  replica matrices, and
the  magnitude  of the  corrections  to  this  approximation were  not
assessed.  They themselves  suggested that  the  calculations actually
done  in  this  paper should  be  carried  out  as  a check  of  their
decimation approximation. Our calculations alas provide no support for it.

JY was supported by WCU program through the KOSEF funded by the MEST
(Grant No. R31-2008-000-10057-0).




\end{document}